\documentclass[aps,pra,twocolumn,showpacs,floatfix]{revtex4}

\usepackage{amsmath}
\usepackage{txfonts}
\usepackage{microtype}
\usepackage{graphicx}
\usepackage{color}
\usepackage{ulem}

\def\sout{\bgroup\markoverwith
  {\textcolor{red}{\rule[0.5ex]{2pt}{0.5pt}}}\ULon}

\graphicspath{{figures/}}

\renewcommand*{\vec}[1]{\boldsymbol{#1}}
\usepackage{slashed}
\def\be{\begin{equation}}
\def\ee{\end{equation}}
\def\bes{\begin{equation*}}
\def\ees{\end{equation*}}
\def\bea{\begin{eqnarray}}
\def\eea{\end{eqnarray}}
\def\beas{\begin{eqnarray*}}
\def\eeas{\end{eqnarray*}}
\def\bal#1\eal{\begin{align}#1\end{align}}
\def\bals#1\eals{\begin{align*}#1\end{align*}}

\newcommand{\ket}[1]{|#1\rangle}

\renewcommand{\vec}[1]{\mathbf{#1}} 
\newcommand{\del}{\partial}

\begin{document}

\title{Above-threshold ionization with highly-charged ions in super-strong laser fields:\\ II.  Relativistic Coulomb-corrected strong field approximation}

\author{Michael \surname{Klaiber}}\thanks{\mbox{Corresponding author: klaiber@mpi-hd.mpg.de }}
\affiliation{Max-Planck-Institut f\"ur Kernphysik, Saupfercheckweg 1, D-69117 Heidelberg, Germany}
\author{Enderalp \surname{ Yakaboylu}} 
\affiliation{Max-Planck-Institut f\"ur Kernphysik, Saupfercheckweg 1, D-69117 Heidelberg, Germany}
\author{Karen Z. \surname{Hatsagortsyan}} 
\affiliation{Max-Planck-Institut f\"ur Kernphysik, Saupfercheckweg 1, D-69117 Heidelberg, Germany}

\date{\today}

\begin{abstract}

We develop a relativistic Coulomb-corrected strong field approximation (SFA) for the investigation of spin effects at above-threshold ionization in relativistically strong laser fields with highly charged hydrogen-like ions.
The Coulomb-corrected SFA is based on the relativistic eikonal-Volkov wave function describing the ionized electron laser-driven continuum dynamics disturbed by the Coulomb field of the ionic core. The SFA in different partitions of the total Hamiltonian is considered. The formalism is applied for direct ionization of a hydrogen-like system in a strong linearly polarized laser field. The differential and total ionization rates are calculated analytically. The relativistic analogue of the Perelomov-Popov-Terent'ev ionization rate is retrieved within the SFA technique. The physical relevance of the SFA in different partitions is discussed.

\end{abstract}


\maketitle

\section{Introduction}

Since the pioneering experiment by Moore et al. \cite{Moore_1999} laser fields of relativistic intensities (exceeding $10^{18}$ W/cm$^2$ at an infrared wavelength) have been applied for the investigation of strong field ionization dynamics of highly charged ions  \cite{Chowdhury_2001,Dammasch_2001,Yamakawa_2003,Gubbini_2005,DiChiara_2008,Palaniyappan_2008,DiChiara_2010}, see also \cite{RMP_2012}. A lot of effort has been devoted to the numerical investigation of the dynamics of highly-charged ions in a super-strong fields \cite{Walser_1999,Hu_1999,Hu_1999b,Casu_2000,Hu_2001,Hu_2002,Walser_2002,Mocken_2004b,Mocken_2008,Hetzheim_2009,Bauke_2011}.

A common analytical approach for strong field atomic processes is the strong field approximation (SFA) \cite{Keldysh_1965,Faisal_1973,Reiss_1980} which has been applied for the treatment of relativistic effects \cite{Reiss_1990,Reiss_1990b}. The main deficiency of the standard SFA is that the influence of the Coulomb field of the atomic core is neglected for the electron dynamics in the continuum and the latter is described  by the Volkov wave function \cite{Volkov_1935}. While this is well-justified for the ionization of a negative ion, for atoms and, moreover, for highly-charged ions it is not valid and the standard SFA can provide only qualitatively correct results. A modification of the theory is required to take into account the influence of the atomic core on the free electron motion.

The Coulomb field effects during ionization have been treated in the Perelomov-Popov-Terent'ev (PPT) theory based on the imaginary time method ~\cite{Perelomov_1967a,Ammosov_1986}. In the PPT theory the barrier formed by the laser and atomic field  is assumed to be quasi-static and the tunneling through it is calculated using the WKB-approximation. The relativistic PPT-theory \cite{Popov_1997,Mur_1998,Milosevic_2002r1,Milosevic_2002r2} has provided the total tunneling rate. However, spin effects are not investigated thoroughly \cite{Popov_2004}.

The standard SFA technique has been modified to include the Coulomb field effects of the atomic core. A heuristic Coulomb-Volkov ansatz has been used for this purpose, see \cite{Jain_Tzoar_1978,Cavaliere_1980,Kaminski_1986,Kaminski_1988,Krstic_1991,Kaminski_1996a,Kaminski_1996b,Ciappina_2007,Yudin_2006,Yudin_2007,Yudin_2008,Faisal_2008}. In a more rigorous way, the SFA is modified by replacing the Volkov wave-function in the SFA transition matrix element by the, so-called, eikonal-Volkov one \cite{Gersten_1975,Krainov_1997}. The latter describes the electron continuum dynamics in the laser and the Coulomb field. Here the laser field is taken into account exactly, while the Coulomb field via the eikonal approximation. The
nonrelativistic Coulomb-corrected SFA based on the eikonal-Volkov wave function  has been applied recently for molecular high-order harmonic generation \cite{Smirnova_2006b,Smirnova_2007,Smirnova_2008}.
The eikonal approximation has been generalized  to include quantum recoil effects \cite{Avetissian_1997,Avetissian_1999}. The  relativistic SFA based on the  generalized eikonal wave function has been proposed in \cite{Avetissian_2001}. However, the final results have been obtained only in the Born approximation.

In this paper we develop the Coulomb corrected SFA for the relativistic regime based on the Dirac equation, extending the corresponding nonrelativistic theory, see paper I (the first paper of this sequel),
into the relativistic domain. This will allow us to calculate spin-resolved  ionization probabilities taking into account accurately the Coulomb field effects of the ionic core. In the Coulomb corrected SFA, the eikonal-Volkov wave function is employed for the description of the final state instead of the Volkov one.  The influence of  the Coulomb potential of the atomic core on the ionized electron continuum dynamics is taken into account via the eikonal approximation. Direct ionization of a hydrogen-like system in a strong linearly polarized laser field is considered. Two versions of the relativistic Coulomb corrected SFA are proposed that are based on the usage of different partitions of the total Hamiltonian in the SFA formalism. The physical relevance of the two versions is discussed.

The plan of the paper is the following: In Sec.~\ref{RCCSFA} the relativistic Coulomb-corrected SFA is developed. The differential and total ionization rates for hydrogen-like systems are derived. The modified SFA using a specific partition of the Hamiltonian is considered in Sec. \ref{modified} and is used for the calculation of ionization rates.

\section{Relativistic Coulomb-corrected SFA}\label{RCCSFA}

We consider the interaction of a highly-charge ion with a strong laser field in the relativistic regime which is described by the time-dependent Dirac equation:
\begin{eqnarray}
\left(i \gamma^\mu \del_\mu + \dfrac{1}{c} \gamma^\mu A_\mu - \gamma^0 \dfrac{V^{(c)}}{c}-c \right) \psi = 0,
\label{Dirac}
\end{eqnarray}
where $\gamma^\mu$ are the Dirac matrices, $V^{(c)}=-\kappa/r$ is the Coulomb potential of the atomic core, $\kappa$ the typical electron momentum in the bound state, determined by the bound state energy via $c^2-I_p=\sqrt{c^4-c^2\kappa^2}$, $I_p$ the ionization potential, $A_\mu$ is the 4-vector potential of the laser field (atomic units are used throughout).
The transition amplitude for the laser induced ionization process from the initial bound state $\phi^{(c)}_{i}$ into an exact continuum state $\psi$ with an asymptotic momentum $\mathbf{p}$ can be written~\cite{Reiss_1990}
\begin{eqnarray}
  M^{(c)}_{fi}=-\dfrac{i}{c} \int d^4 x \overline{\psi}(\mathbf{r},t)\slashed{A}\phi^{(c)}_{i}(\mathbf{r},t),
  \label{M_rel}
\end{eqnarray}
with $\slashed{A}\equiv \gamma^\mu A_\mu$. Equation~(\ref{M_rel}) is still exact with the exact wave function $\psi(\mathbf{r},t)$ describing the dynamics of the electron in the ionic and the laser field. Here, we have used the  standard partition of the total Hamiltonian:
\begin{eqnarray}
  H\,\,\,&=&H_0+H_{int},\\
H_0\,\, &=& c\boldsymbol{\alpha}\cdot \hat{\mathbf{p}}   +\beta c^2+V^{(c)}(r)\\
H_{int} &=& \beta\slashed{A},
\end{eqnarray}
where $\boldsymbol{\alpha} = \gamma^0 \boldsymbol{\gamma}$, $\beta=\gamma^0$ are the Dirac matrices, and $c$ the speed of light.
We consider a hydrogen-like highly charged ion in the relativistic parameter regime, i.e., the wave function of the initial bound state $\phi^{(c)}_{i}$ fulfills the Dirac equation with the Hamiltonian $H_0$ and
is  given by~\cite{Bjorken_1964}
\begin{eqnarray}
  \phi_{i}^{(c)}(\mathbf{r},t)&=&\frac{\kappa^{3/2}}{\sqrt{\pi}}\sqrt{\frac{2-\frac{I_p}{c^2}}{\Gamma\left(3-\frac{2I_p}{c^2}\right)}}(2\kappa r)^{-\frac{I_p}{c^2}}\nonumber\\
  &&\times\exp\left[-\kappa r-i(c^2-I_p)t\right]v_i,
  \label{gs}
\end{eqnarray}
where the bispinor 
\begin{eqnarray}
  v_i=\left(\begin{array}{c}\chi_i \\
i\frac{I_p}{c\kappa}\frac{\boldsymbol{\sigma}\cdot\mathbf{r}}{r}\chi_i\end{array}\right)
\end{eqnarray}
describes the spin-up and -down states ($i=\pm$), with the two-component spinors  $\chi_+=(1,0)$ and $\chi_-=(0,1)$, respectively, and the Pauli-matrices $\boldsymbol{\sigma}$.

The hydrogen-like system interacts with a strong linearly polarized laser field
\begin{eqnarray}
  \mathbf{E}(\eta)=-\mathbf{E}_0\cos(\omega\eta),
\end{eqnarray}
where $\eta=k^\mu x_\mu/\omega$, and $k^\mu=(\omega/c,\mathbf{k})$ is the laser 4-wave-vector. 
[the coordinate and the momentum projections are defined as $r_E\equiv \textbf{r}\cdot \hat{\textbf{e}}$, $r_k\equiv \textbf{r}\cdot \hat{\textbf{k}}$, $p_E\equiv \textbf{p}\cdot \hat{\textbf{e}}$, $p_k\equiv \textbf{p}\cdot \hat{\textbf{k}}$, and $p_B\equiv \textbf{p}\cdot (\hat{\textbf{k}}\times \hat{\textbf{e}})$, with the unit vectors $\hat{\mathbf{k}}$ and $\hat{\mathbf{e}}$ in the laser propagation  and the polarization direction, respectively].
The vector-potential is chosen in the G{\"o}ppert-Mayer gauge:
\begin{eqnarray}
A^{\mu}\equiv (\varphi,c\mathbf{A})=(-\mathbf{r}\cdot\mathbf{E},-\hat{\mathbf{k}}\,(\mathbf{r}\cdot\mathbf{E})).
\label{A_GM} 
\end{eqnarray}
Generally, the SFA in different gauges do not coincide and correspond to different physical approximations \cite{Faisal_2007a,Faisal_2007b,Bauer_2005}. In paper I we have seen that in the nonrelativistic regime the Coulomb-corrected SFA in the length gauge leads, first, to rather simple expressions for the Coulomb-corrected ionization  amplitude, see Eq. (I.28) [refers to Eq. (28) of paper I], which is due to the cancellation of the $\textbf{r}\cdot \textbf{E}$ interaction Hamiltonian in the matrix element by the Coulomb-correction factor; and, second, to results coinciding with the PPT-ionization rates. As the latter provides a good approximation to experimental observations, in the relativistic regime we choose a gauge which generalizes the length gauge into the relativistic domain \cite{Reiss_1992}, that is the G{\"o}ppert-Mayer gauge defined by Eq. (\ref{A_GM}).

In the conventional SFA the exact continuum state is approximated by the Volkov-wavefunction which is identical to the first order WKB-approximation of the continuum electron in the laser field. A systematic improvement of this approximation is achieved in the relativistic Coulomb-corrected SFA by the replacement of the exact wave function $\psi(\mathbf{r},t)$ with the relativistic eikonal-Volkov wave function. Similar to the non-relativistic case, see paper I, first we apply the WKB approximation to the solution of the Dirac equation (\ref{Dirac}), then the resulting equations will be solved via a perturbative expansion in the Coulomb potential $V^{(c)}$.

Due to the gauge covariance of the Dirac equation, we switch to the velocity gauge with its vector potential $\tilde{A}^\mu = \tilde{A}^\mu(\eta) = (0, c\tilde{\vec{A}})$, $\tilde{\mathbf{A}}=-\int^\eta_{-\infty} d\eta'\mathbf{E}(\eta')$ and after solving the wave-equation go back to the G{\"o}ppert-Mayer gauge. The quadratic Dirac equation in velocity gauge becomes
\begin{widetext}
\bea
&& \left(i \hbar \gamma^\mu \del_\mu + \dfrac{1}{c} \gamma^\mu \tilde{A}_\mu - \gamma^0 \dfrac{V^{(c)}}{c}+c \right)\left(i \hbar \gamma^\mu \del_\mu + \dfrac{1}{c} \gamma^\mu \tilde{A}_\mu - \gamma^0 \dfrac{V^{(c)}}{c}-c \right) \psi = 0, \\
&& \left[-\hbar^2 \del^2 + \dfrac{i \hbar}{c} \left( \slashed{k} \slashed{\tilde{A}'} + 2 \tilde{A} \cdot \del + \boldsymbol{\alpha} \cdot \boldsymbol{\nabla} V^{(c)} -2 V^{(c)} \del_0 \right) + \dfrac{\tilde{A}^2}{c^2} + \dfrac{{V^{(c)}}^2}{c^2} -c^2 \right] \psi = 0.
\eea
where $\tilde{A}'\equiv \frac{1}{\omega}\frac{d\tilde{A}}{ d\eta }$, and we have inserted $\hbar$ to indicate the WKB-expansion.
Let us assume that the solution has the form $\psi= e^{i S / \hbar}$, then the corresponding equation for $S$ is
\bea
\nonumber  \left(\del_\mu S - \dfrac{\tilde{A}_\mu}{c} + g_{\mu 0}\dfrac{V^{(c)}}{c}\right)\left(\del^\mu S - \dfrac{\tilde{A}^\mu}{c} + g^{\mu 0}\dfrac{V^{(c)}}{c}\right) 
+ \dfrac{\hbar}{i} \left(\del^2 S- \dfrac{\slashed{k} \slashed{\tilde{A}'}}{c} - \dfrac{\boldsymbol{\alpha}\cdot \boldsymbol{\nabla} V^{(c)}}{c} \right) = c^2.
\eea
The WKB expansion $S=S_0 + \dfrac{\hbar}{i} S_1 + \ldots$ yields following equations up the first order in $\hbar / i$
\bea
&\left(\dfrac{\hbar}{i}\right)^0 :& \quad \del_\mu S_0 \del^\mu S_0 - \dfrac{2 \tilde{A}_\mu \del^\mu S_0}{c} + \dfrac{\tilde{A}^2}{c^2}  -c^2 = -\dfrac{2V^{(c)} \del_0 S_0}{c}- \dfrac{{V^{(c)}}^2}{c^2} , \\
&\left(\dfrac{\hbar}{i}\right)^1 :&  2 \del_\mu S_0 \del^\mu S_1 -\dfrac{2 \tilde{A}_\mu \del^\mu S_1}{c}= -\del^2 S_0 + \dfrac{\slashed{k} \slashed{\tilde{A}'}}{c}-\dfrac{2 V^{(c)} \del_0 S_1}{c}  + \dfrac{\boldsymbol{\alpha}\cdot \boldsymbol{\nabla} V^{(c)}}{c}.
\eea
\end{widetext}
These equations can be solved perturbatively with respect to the potential term $V^{(c)}$. If we define
\bea
S_0 &=& S^{(0)}_0 + S^{(1)}_0, \label{expansion1}\\
S_1 &=& S^{(0)}_1 + S^{(1)}_1,
\label{expansion2}
\eea
the following equations can be derived
\bea
\label{vol_act}&& \del_\mu S^{(0)}_0 \del^\mu S^{(0)}_0-\dfrac{2 \tilde{A}_\mu \del^\mu S^{(0)}_0}{c} + \dfrac{\tilde{A}^2}{c^2} - c^2 =0, \\
\label{coul_corr_act}&& \del_\mu S^{(1)}_0 \pi^\mu =  -\dfrac{ V^{(c)}}{c}\pi^0, \\
\label{vol_spi}&& 2 \del_\mu S^{(0)}_1 \pi^\mu = \del^2 S^{(0)}_0  - \slashed{k} \slashed{\tilde{A}'}, \\
\nonumber && \del_\mu S^{(1)}_1\pi^\mu = -\dfrac{\boldsymbol{\alpha}\cdot (\boldsymbol{\nabla} V^{(c)})}{2 c} \\
\label{coul_corr_spi} &&+ \dfrac{\del^2 S^{(1)}_0}{2} + \del_\mu S^{(1)}_0 \del^\mu S^{(0)}_1+ \dfrac{V^{(c)} \del_t S^{(0)}_1}{c^2},
\eea
where $\pi^\mu = -\del^\mu S^{(0)}_0 + \dfrac{\tilde{A}^\mu}{c}$ is the relativistic momentum. Eq. (\ref{vol_act}) gives the Volkov-action, i.e.,
\be
\label{vol_act_sol}
S^{(0)}_0 = - p \cdot x - \dfrac{\omega}{ c\,(k\cdot p)} \int^\infty_\eta \left(\tilde{A} \cdot p + \dfrac{\tilde{A}^2}{2 c}\right) d \eta'.
\ee
On the other hand, the solution of Eq. (\ref{vol_spi}) is
\be
S^{(0)}_1 = -\dfrac{\slashed{k} \slashed{\tilde{A}}}{2c\,( k \cdot p)}.
\ee
Here we should note that $S^{(0)}_0$ and $S^{(0)}_1$ generate the full Volkov solution. 
The corresponding equations in the G{\"o}ppert-Mayer gauge can be found via a gauge transformation with an generating function $\chi= \tilde{\mathbf{A}}(\eta) \cdot \vec{r} $. After a coordinate transformation from $(t, \vec{r})$ to $(\eta, \vec{r})$ they become
\begin{eqnarray}
  S^{(0)}_0(\mathbf{r},\eta)&=&\left(\mathbf{p}+\tilde{\mathbf{A}}(\eta)-\frac{\varepsilon}{c}\hat{\mathbf{k}}\right)\cdot\mathbf{r}\\
  &&+\int^{\infty}_{\eta}\eta'\left( \varepsilon +\frac{(\mathbf{p}+\tilde{\mathbf{A}}(\eta')/2)\cdot\tilde{\mathbf{A}}(\eta')}{\Lambda}\right)\nonumber, \\
S^{(0)}_1 (\eta) &=& \dfrac{(1+ \hat{\vec{k}}\cdot \boldsymbol{\alpha}) \,( \boldsymbol{\alpha}\cdot \tilde{\mathbf{A}})}{2 c \Lambda}, 
\end{eqnarray}
 with  $\varepsilon=c \sqrt{c^2+\mathbf{p}^2}$ and the constant of motion $\Lambda\equiv k \cdot p/\omega$ 
Further, Eqs.~(\ref{coul_corr_act}) and (\ref{coul_corr_spi}) can be solved via the method of characteristics as
\bea
\label{coul_corr_act_eq} \dfrac{d S^{(1)}_0}{d \sigma} &=& \del_\mu S^{(1)}_0 \dfrac{d x^\mu}{d \sigma} =- \dfrac{ V^{(c)}}{c}\pi^0, \\
\label{coul_corr_spi_eq} \dfrac{d S^{(1)}_1}{d \sigma} &=& \del_\mu S^{(1)}_1 \dfrac{d x^\mu}{d \sigma} \\
\nonumber &=& -\dfrac{\boldsymbol{\alpha}\cdot \boldsymbol{\nabla} V^{(c)}}{2 c} + \dfrac{\del^2 S^{(1)}_0}{2} + \del_\mu S^{(1)}_0 \del^\mu S^{(0)}_1+ \dfrac{V^{(c)} \del_t S^{(0)}_1}{c^2},
\eea
where the trajectory is given by
\be
\dfrac{d x^\mu}{d \sigma} = \pi^\mu.
\ee
This leads to $d \sigma = d \eta/\Lambda$,
with the relativistic kinetic momentum in the laser field
\begin{eqnarray}
  \boldsymbol{\pi}(\eta)&\equiv &\mathbf{p}(\eta) =\boldsymbol{\nabla}S^{(0)}_0+\tilde{\mathbf{A}}(\eta)\nonumber\\
  &=&\mathbf{p}+\tilde{\mathbf{A}}(\eta)+\hat{\mathbf{k}}\frac{(\mathbf{p}+\tilde{\mathbf{A}}(\eta)/2)\cdot\tilde{\mathbf{A}}(\eta)}{c\Lambda},
\end{eqnarray}
and the corresponding relativistic kinetic energy
\begin{eqnarray}
 c \pi^0 (\eta)\equiv \varepsilon(\eta)=-\partial_t{S}^{(0)}_0 
  =\varepsilon+\frac{(\mathbf{p}+\tilde{\mathbf{A}}(\eta)/2)\cdot\tilde{\mathbf{A}}(\eta)}{\Lambda}.
\end{eqnarray}
Then the solutions of Eqs. (\ref{coul_corr_act_eq}) and (\ref{coul_corr_spi_eq}) read 
\bea
\label{coul_corr_act_sol} S^{(1)}_0(\mathbf{r},\eta) &=&\int^{\infty}_{\eta} d\eta'\frac{\varepsilon(\eta')}{c^2\Lambda}V^{(c)}\left(\mathbf{r}(\eta')\right),\\
\label{coul_corr_spi_sol} S^{(1)}_1(\mathbf{r},\eta) &=& \int_\eta^\infty  \dfrac{d\eta'}{ \Lambda}\left[\dfrac{\boldsymbol{\alpha}\cdot \boldsymbol{\nabla} V^{(c)}\left(\mathbf{r}(\eta')\right)}{2 c} \right. \\
\nonumber &-& \left.  \dfrac{V^{(c)}\left(\mathbf{r}(\eta')\right) \del_0 S^{(0)}_1}{c}-\del_\mu S^{(1)}_0 \del^\mu S^{(0)}_1- \dfrac{\del^2 S^{(1)}_0}{2} \right]
\eea
with the relativistic trajectory of the electron in the laser field $\mathbf{r}(\eta')=\mathbf{r}+\int^{\eta'}_{\eta}d\eta''\mathbf{p}(\eta'')/\Lambda$, starting at the ionization phase $\eta$ with coordinate $\mathbf{r}$.

We can evaluate the explicit parameters for the applied eikonal approximation estimating the imposed conditions for the expansion of Eqs. (\ref{expansion1}) and (\ref{expansion2}): $S_0^{(1)}\ll S_0^{(0)}$, and $S_1^{(0)}\ll S_0^{(1)}$. 
For this purpose we use order of magnitude estimations:
\begin{eqnarray}
S_0^{(0)}&\sim& I_p\tau_c\sim \frac{E_a}{E_0},\label{S00}\\
S^{(1)}_0 &\sim& \log\left(\sqrt{\frac{E_a}{E_0}}\right) \sim 1\label{S10}\\
S^{(0)}_1 &\sim& \frac{\tilde{A}}{c \Lambda}\sim \frac{E_0\tau_c}{c}\sim \sqrt{\frac{I_p}{c^2}} \label{S01}\\
S^{(1)}_1 &\sim& \int dt \frac{\dot{r}_E(t)}{cr_E(t)^2}\sim\frac{1}{cr_{E\,c}}\sim\sqrt{\frac{I_p}{c^2}}\sqrt{\frac{E_0}{E_a}}.\label{S11}
\end{eqnarray}
Here, the first two equations were already derived in paper I, see Eq. (I.22), $E_a\equiv (2I_p)^{3/2}$.  Further, taking into account the typical time during tunneling $\tau_c\sim \gamma/\omega=\kappa/E_0$, with the Keldysh parameter $\gamma=\omega\kappa/E_0$ \cite{Keldysh_1965}, the value of the typical velocity of the electron during tunneling $\dot{r}_{E\,c}\sim\kappa$ [$\kappa\approx \sqrt{2I_p}]$, the interaction length $r_{E\,c}\sim\dot{r}_{E\,c}\delta\tau_c$, the uncertainty of the initial time $\delta\tau_c\sim   1/\sqrt{\kappa E_0}$ [from the saddle-point integration $\delta\tau_c^2\sim 1/\ddot{\tilde{S}}(t_s)$], $\varepsilon (\eta)_c\sim c^2$, $\Lambda \sim 1$ [the electron is at rest at the tunnel exit], and $\tilde{A}\sim E_0\tau_c$, we come to the following conditions
for the applicability of the eikonal approximation: 
\begin{eqnarray}
\frac{E}{E_a} \ll 1\,\,\,\,\,\,\,\,\,{\rm and} \,\,\,\,\,\,\,\,\,\sqrt{\frac{I_p}{c^2}}\ll 1.
\end{eqnarray}
The action function $S_1^{(1)}$ can be approximated by the first term in Eq. (\ref{coul_corr_spi_sol}):
\be
 S^{(1)}_1(\mathbf{r},\eta) = \int_\eta^\infty  d\eta'\dfrac{\boldsymbol{\alpha}\cdot \boldsymbol{\nabla} V^{(c)}\left(\mathbf{r}(\eta')\right)}{2 c\Lambda}.
\ee
In fact, the order of magnitude of the terms in r.h.s. of Eq.~(\ref{coul_corr_spi_sol}) are 
\begin{eqnarray}
{\cal T}_1&\sim& \int d\eta\frac{V^{(c)}}{c r_{E\,c}}\sim \sqrt{\frac{E_0}{E_a}}\sqrt{\frac{I_p}{c^2}},\label{T1}\\
{\cal T}_2&\sim& \frac{E_0}{c^3}\int  d\eta V^{(c)}\sim\frac{E_0}{E_a}\left(\frac{I_p}{c^2}\right)^{3/2},\\
{\cal T}_3&\sim& \frac{E_0}{c^2}\int   d\eta V^{(c)}\frac{\tau_c}{r_{E\,c}}\sim\frac{E_0}{E_a}\frac{I_p}{c^2},
\end{eqnarray}
and ${\cal T}_4$ is vanishing because in the tunneling region $\Delta V^{(c)}=0$. From the latter, we estimate the ratios:
\begin{eqnarray}
\frac{{\cal T}_2}{{\cal T}_1}&\sim& \frac{I_p}{c^2}\sqrt{\frac{E}{E_a}}\ll 1,\\
\frac{{\cal T}_3}{{\cal T}_1}&\sim& \sqrt{\frac{I_p}{c^2}}\sqrt{\frac{E}{E_a}}\ll 1,
\end{eqnarray}
therefore, only the first integrand term may be maintained in Eq.~(\ref{coul_corr_spi_sol}).
Further, one can see from Eqs. (\ref{S01}) and (\ref{S11}) that $S^{(0)}_1\ll 1$ and $S^{(1)}_1\ll 1$, which allows us to expand  the corresponding exponents.

Thus, the relativistic eikonal-Volkov wave function reads
\begin{eqnarray}
\nonumber  && \psi_f(\mathbf{r},t) \\
&=&\left[1+\frac{(1+\hat{\mathbf{k}}\cdot\boldsymbol{\alpha})\,\boldsymbol{\alpha}\cdot\tilde{\mathbf{A}}(\eta)+\int_\eta^\infty  d\eta'\boldsymbol{\alpha}\cdot \boldsymbol{\nabla} V^{(c)}\left(\mathbf{r}(\eta')\right)}{2c\Lambda}\right]\nonumber\\
  &&\times \frac{c \,u_f}{\sqrt{(2\pi)^{3}\varepsilon}}\exp[i S^{(0)}_0 (\mathbf{r},\eta)+i S^{(1)}_0(\mathbf{r},\eta)] ,
\label{vrel}
\end{eqnarray}
with the bispinor $u_f$ for the spin-up and -down final states ($f=\pm$)
\begin{eqnarray}
  u_f=\left(\begin{array}{c}\sqrt{(1+\varepsilon/c^2)/2}\,\chi_f\\ \frac{(\boldsymbol{\sigma}\cdot\mathbf{p})\,\chi_f}{\sqrt{2(c^2+\varepsilon)}}\end{array}\right),
\end{eqnarray}
where $\chi_+=(1,0)$ and $\chi_-=(0,1)$. 
The last term in the pre-exponential [$\propto {\nabla}V^{(c)}$] describes the spin-orbit coupling during ionization (under-the-barrier motion). Via a $p/c$-expansion of the Dirac equation in the atomic and the laser field it can be shown that the spin-orbit coupling, given in the Hamiltonian by the term
\begin{equation}
 H_{SO}=\frac{\boldsymbol{\sigma}\cdot\left[\mathbf{p}(\eta)\times\boldsymbol{\nabla}V^{(c)}(\mathbf{r}(\eta))\right]}{4c^2},
\end{equation}
can be evaluated along the most probable trajectory as follows
\begin{equation}
 H_{SO}\tau_c\sim \frac{p_k }{c^2}\frac{\partial V^{(c)}}{\partial r_E}\tau_c,\,\frac{p_E }{c^2}\frac{\partial V^{(c)}}{\partial r_k}\tau_c\sim \left(\frac{I_p}{c^2}\right)^{3/2},
\end{equation}
where the typical values for $p_E\sim \kappa$, $p_k\sim \kappa^2/c$, $r_k\sim r_{E\,c}\kappa/c$ have been used. 
It is of higher order smallness than the terms kept in the eikonal wave function
and is neglected in the further calculation.

Finally, the ionization amplitude in the Coulomb-corrected SFA will have the following form in the relativistic regime:
\begin{eqnarray}
  M^{(c)}_{fi}&=&-i\int^{\infty}_{-\infty} d\eta\langle\mathbf{p}(\eta)_r,m|H_{int}\exp[-i S^{(1)}_0(\mathbf{r},\eta))]|0_r,i\rangle\nonumber\\
  &&\times\exp[-i\tilde{S}(\eta)].
  \label{M_rel1}
\end{eqnarray}  
with the spatial parts of the initial  $|0_r,i\rangle$ and the final $|\mathbf{p}(\eta)_r,f\rangle$ spinors and the contracted action
\begin{eqnarray}
  \tilde{S}(\eta)=\int^{\infty}_{\eta}d\eta'\left\{\varepsilon-c^2+I_p+\frac{[\mathbf{p}+\tilde{\mathbf{A}}(\eta')/2]\cdot\tilde{\mathbf{A}}(\eta')}{\Lambda}\right\}.
\end{eqnarray}

In the adiabatic regime ($\omega\ll I_p, U_p$, with the ponderomotive potential $U_p=E^2_0/4\omega^2$) the time integration in the amplitude of Eq.~(\ref{M_rel1})  is calculated with SPM. As in the nonrelativistic case, here also the disturbance of the saddle-point conditions due to the Coulomb field is neglected. This is in accordance with our approach to take into account the Coulomb field perturbatively in the phase of the WKB wave function: $S^{(1)}_{0}$ is already proportional to the Coulomb field $V^{(c)}(r)$, therefore, in the trajectory $\mathbf{r}(\eta)$ it should be neglected. Mathematically this is justified as one can see from  the estimation $\partial_\eta S^{(1)}_0\sim I_p\sqrt{E_0/E_a}$ [similar to estimations in Eqs. (\ref{S00})-(\ref{S11})]. It shows that 
$\partial_{\eta}S^{(1)}$ is negligibly small compared to $I_p$, and, consequently, the saddle-point condition $\partial_\eta \tilde{S}(\eta_s)=0$ is not disturbed.
The latter leads to the following two saddle points per cycle:
\begin{eqnarray}
  \omega\eta_1&=&\arcsin\left[-\frac{p_E}{E_0/\omega}+i\sqrt{\frac{2\Lambda(\varepsilon-c^2+I_p)-p_E^2}{(E_0/\omega)^2}}\right]\\
  \omega\eta_2&=&\pi-\arcsin\left[-\frac{p_E}{E_0/\omega}-i\sqrt{\frac{2\Lambda(\varepsilon-c^2+I_p)-p_E^2}{(E_0/\omega)^2}}\right].\nonumber
\end{eqnarray}
The amplitude is now evaluated at these phases.
First, we consider the action $\tilde{S}(\eta_s)$ in the exponent:
\begin{eqnarray}
  {\rm Im}\,\{\tilde{S}(\eta_s)\}=\frac{\left[2\Lambda(\varepsilon-c^2+I_p)-p_E^2\right]^{3/2}}{3|\mathbf{E}(\eta_0)|\Lambda}
\end{eqnarray}
with $|\mathbf{E}(\eta_0)|=E_0\sqrt{1-(p_E/(E_0/\omega))^2}$.
Since the real part gives  an unimportant phase in the resulting amplitude, only the imaginary part is given. This function in the exponent dominates the momentum distribution and determines the maximum
of the momentum distribution which is located at a parabola with $p_k=I_p/3c+p_E^2/2c(1+I_p/3c^2)$, $p_B=0$. In the following, we, therefore, evaluate the less important pre-exponential factor only at this parabola and neglect deviations from it. In evaluating $S^{(1)}_0$ at the saddle point $\eta=\eta_s$, note that the integration in $S^{(1)}_0$ starts at the saddle point $\eta_s$, when the electron enters the barrier due to the effective potential, and goes to infinity when the electron is far away from the core. However,
the upper integration limit could be set at the moment $\eta_0$, when the electron leaves the barrier since further integration over the continuum  leads eventually only to an unimportant phase of the amplitude.
The time which the electron spends under the barrier is short 
compared with the laser period and the integrand can be expanded around the saddle point $\eta_s$:
\begin{eqnarray}
  &&-i S^{(1)}(\mathbf{r},\eta_s)=\int^{\eta_0}_{\eta_s} d\eta'\frac{i\kappa}{\Lambda c^2}\label{Sexact}\\
  &&\times\frac{\varepsilon(\eta_s)+\partial_{\eta}\varepsilon(\eta_s)(\eta'-\eta_s)+\frac{\partial^2_{\eta,\eta}\varepsilon(\eta_s)}{2}(\eta'-\eta_s)^2}{\left| \mathbf{r}+\frac{\mathbf{p}(\eta_s)}{\Lambda}(\eta'-\eta_s)+\frac{\mathbf{F}_L(\eta_s)}{2\Lambda}(\eta'-\eta_s)^2+\frac{\hat{\mathbf{k}}E(\eta_s)^2}{6c\Lambda^2}(\eta-\eta_s)^3\right| },\nonumber
  \end{eqnarray}
where $\mathbf{F}_L(\eta)=-\mathbf{E}(\eta)-\hat{\mathbf{k}}(\mathbf{p(\eta)}\cdot\mathbf{E}(\eta))/c\Lambda$ is the Lorentz-force due to the laser field. As the correction factor is evaluated at the parabola with $p_k=I_p/3c+p_E^2/2c(1+I_p/3c^2)$ and $p_B=0$,
we derive the trajectories at these special values of momentum. During the motion under the barrier the magnitude of the coordinate variation in the $\hat{\textbf{k}}$-direction (imaginary) is significantly smaller than in  $\hat{\textbf{e}}$-direction [$r_k(\eta)/r_E(\eta)\sim \kappa/c$] and the integrand can be expanded by $ r_k(\eta)$. It is taken into account also that the ionization event happens close to the laser polarization axis: $r_k\ll r_E$. The integral is, therefore, approximated by:  
\begin{eqnarray}
  &&-i S^{(1)}(\mathbf{r},\eta_s)=\int^{\eta_0}_{\eta_s} d\eta'\frac{i\kappa}{\Lambda c^2}\nonumber\\
  &&\times\frac{\varepsilon(\eta_s)+\partial_{\eta}\varepsilon(\eta_s)(\eta'-\eta_s)+\frac{\partial^2_{\eta,\eta}\varepsilon(\eta_s)}{2}(\eta'-\eta_s)^2}{\left| x+\frac{p_E(\eta_s)}{\Lambda}(\eta'-\eta_s)-\frac{E(\eta_s)}{2\Lambda}(\eta'-\eta_s)^2\right| }\nonumber\\
&&-\int^{\eta_0}_{\eta_s} d\eta'\frac{i\kappa}{2} \frac{r_k(\eta')^2}{\left|x+\frac{p_E(\eta_s)}{\Lambda}(\eta'-\eta_s)-\frac{E(\eta_s)}{2\Lambda}(\eta'-\eta_s)^2\right|^3},
 \label{S_1ex2}
\end{eqnarray}
where 
\begin{eqnarray}
  r_k(\eta)&=&\frac{i\sqrt{2I_p}r_E}{3c}-\frac{2 I_p}{3c}(\eta-\eta_s) + \frac{i E(\eta_0)\sqrt{2 I_p}}{2c} (\eta-\eta_s)^2\nonumber\\
  &&- \frac{E(\eta_0)^2}{6c}(\eta-\eta_s)^3,\\
p_E(\eta_s)&=&i\sqrt{2\Lambda(\varepsilon-c^2+I_p)-p_E^2}\approx i\sqrt{2I_p}\left(1 - 5 I_p/36c^2\right).\nonumber \\
\\
\varepsilon(\eta_s)\,\,\,\,&=&c^2-I_p\\
 \partial_{\eta}\varepsilon(\eta_s)\,\,\,&=&-p_E(\eta_s)E(\eta_0)/\Lambda\\
\partial^2_{\eta,\eta}\varepsilon(\eta_s)&= &E(\eta_0)^2/\Lambda
\end{eqnarray}
Other parameters are $\omega\eta_0=\arcsin\left[-p_E/E_0/\omega\right]$, 
$\kappa\approx\sqrt{2I_p}(1- I_p/4c^2)$,
$\Lambda\approx 1-I_p/3c^2$,
$r_B(\eta)\approx 0$,
$E(\eta_s)\approx E(\eta_0)$.
To have compact expressions,  expansions on the small parameter $I_p/c^2$ have been used above [$I_p/c^2\approx 0.25$ for hydrogen-like uranium]. The starting coordinate of the trajectory $\mathbf{r}$ is found from the saddle-point condition of the integral $\int d^3\mathbf{r}\exp[-i\mathbf{p}(\eta_s)\cdot\mathbf{r}-\kappa r]$, that is $\mathbf{r}_s/r_s=\mathbf{p}(\eta_s)/(i\kappa)$.
With these parameters the integral in Eq.~(\ref{S_1ex2}) can be calculated: 
\begin{eqnarray}
  Q_{r}&\equiv&
  \exp\left[-i S^{(1)}(\mathbf{r},\eta_s)\right]\nonumber\\
  &=&\exp\left[\frac{2I_p}{c^2}\right]\frac{1-\frac{I_p}{6c^2}}{\lambda^{1-\frac{I_p}{c^2}}}\nonumber\\
  &=&\exp\left[\frac{2I_p}{c^2}\right]\left(1-\frac{I_p}{6c^2}\right)Q^{1-\frac{I_p}{c^2}}_{nr}
  \label{Qeq}
\end{eqnarray}
with the small parameter $\lambda=-\mathbf{r}\cdot\mathbf{E}(\eta_s)/4I_p$. 
In the expression above only the leading order term in $1/\lambda$ are retained.
\begin{figure}
  \begin{center}
    \includegraphics[width=0.4\textwidth]{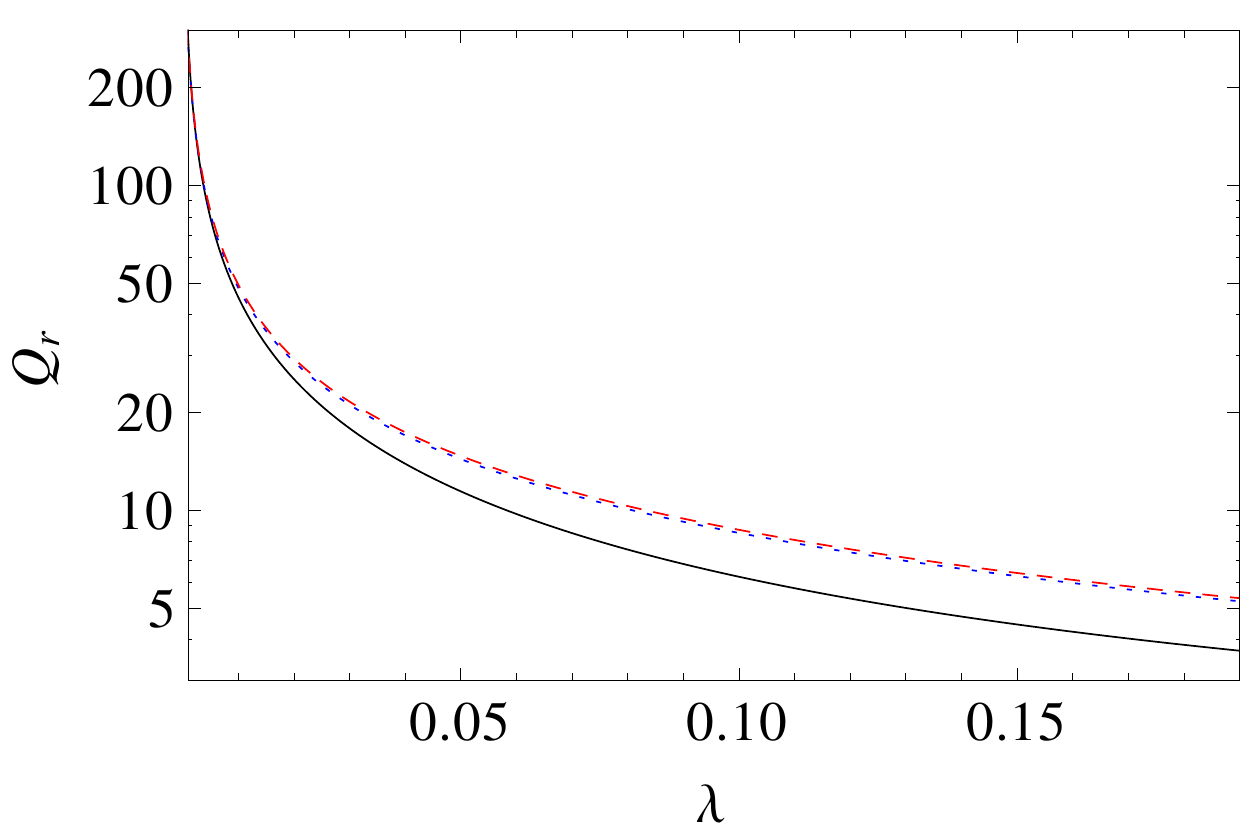}
       \caption{The Coulomb-correction factor $Q_r$ vs the parameter $\lambda$ for $\gamma=0.01$ and $I_p/c^2=0.25$ (equivalent to $Z=91$): (black, solid) the numerical calculation via Eq. (\ref{Sexact}) without approximations, (blue, short-dashed) the analytical result of Eq.~(\ref{Qeq}), (red, long-dashed) the Coulomb-correction factor of the relativistic PPT \cite{Milosevic_2002r1}. }
    \label{Q}
  \end{center}
\end{figure}
For justification of our approximations, we compare in Fig.~\ref{Q} our analytically derived Coulomb-correction factor $Q_{r}$ with the exact result of a numerical calculation of Eq.~(\ref{Sexact}), and with the Coulomb-correction factor of the relativistic PPT~\cite{Milosevic_2002r1}.
Deviation of our approximate Coulomb-correction factor from the exact one occurs mostly due to a linearization of the $\lambda$-dependence in the analytical expressions which, however, is needed for the further analytical integration.

Using the Coulomb-correction factor $Q_{r}$ of Eq. (\ref{Qeq}),  the spatial integration in Eq.~(\ref{M_rel1}) can be carried out. The Coulomb-corrected pre-exponential matrix element reads
\begin{eqnarray}
  \tilde{m}_{fi}(\eta_s)=\langle\mathbf{p}(\eta_s)_r,f|\mathbf{r}\cdot\mathbf{E}(\eta_s)(1-\hat{\mathbf{k}}\cdot\boldsymbol{\alpha})Q_r|0_{r},i\rangle. 
  \label{me_rel}
\end{eqnarray}
It is remarkable that the Coulomb-correction factor in Eq. (\ref{me_rel}) cancels the dependence of the matrix element on the electric dipole operator $\mathbf{r}\cdot\mathbf{E}$ and, approximately, also the pre-exponential term $r^{-I_p/c^2}$ of the initial state wave function [when the approximation $r\approx r_E$, i.e., ionization happens at the laser polarization axis, is used]. This is a consequence of the applied G\"oppert-Mayer gauge which significantly simplifies the  calculation of the ionization matrix element.

In this paper we are concerned with spin-unresolved ionization probabilities. Therefore, we are free to choose the orientation of the quantization axis. In the following we assume that the initial spinor of the bound state is aligned  along the laser magnetic field direction. The spin-dependent matrix element in Eq.~(\ref{me_rel}) in this case yields:
\begin{eqnarray}
  \tilde{m}_{fi}(\eta_s)&=&\int d^{3}\mathbf{r}\frac{2c\kappa^{3/2}I_p}{\pi^2\sqrt{2\varepsilon}}\sqrt{\frac{2-\frac{I_p}{c^2}}{\Gamma(3-\frac{2I_p}{c^2})}}\left(\frac{8\kappa I_p}{|E(\eta_s)|}\right)^{-\frac{I_p}{c^2}}\nonumber\\
  &&\times\exp\left[\frac{2I_p}{c^2}\right]\exp\left[-i\mathbf{p}(\eta_s)\cdot\mathbf{r}-\kappa r\right]\label{m_fi11}\\
  &&\times\left(1-\frac{I_p}{6c^2}\right)u^+_f\left(1-\hat{\mathbf{k}}\cdot\boldsymbol{\alpha}\right)v_i\nonumber\label{m_fi1}\\
  &=&\frac{2^{3-\frac{2I_p}{c^2}}(2I_p)^{\frac{9}{4}-\frac{3I_p}{2c^2}}|E(\eta_s)|^{\frac{I_p}{c^2}}\exp\left[\frac{2I_p}{c^2}\right]}{\pi\sqrt{\Gamma\left(3-\frac{2I_p}{c^2}\right)}\left[\mathbf{p}(\eta_s)^2+\kappa^2\right]^2}\,m_{fi},
\label{m_fi2}
\end{eqnarray}
where the factor $m_{fi}$ for different spin transitions is
\begin{eqnarray}
m_{++}&=&\frac{\sqrt{2}c(2 c+i p_E)\left(1+\frac{\beta}{2}\right)}{ \sqrt{8 c^4+6 c^2 p_E^2+p_E^4}}\label{m_++}\\
&-&\frac{\beta ^2 c \left(168 c^4-16 i c^3 p_E+142 c^2 p_E^2-8 i c p_E^3+25 p_E^4\right)}{24
   \sqrt{2} (2 c-i p_E) \left(2 c^2+p_E^2\right)^{3/2} \sqrt{4 c^2+p_E^2}},\nonumber\\
m_{--}&=&\frac{\sqrt{2}c(2 c-i p_E)\left(1-\frac{\beta}{2}\right)}{ \sqrt{8 c^4+6 c^2 p_E^2+p_E^4}}\label{m_--}\\
&-&\frac{\beta ^2 c \left(168 c^4+16 i c^3 p_E+142 c^2 p_E^2+8 i c p_E^3+25 p_E^4\right)}{24
   \sqrt{2} (2 c+i p_E) \left(2 c^2+p_E^2\right)^{3/2} \sqrt{4 c^2+p_E^2}},\nonumber
\end{eqnarray}
with $\beta=\sqrt{2I_p}/c$ and $m_{-+}=m_{+-}=0$.
Eqs. (\ref{m_++}) and (\ref{m_--}) are exact in $p_E/c$, the $p_E/c$ term can be important at $\xi\gg 1$.
As in the non-relativistic case, see paper I, the matrix element has a singularity at the saddle point $\eta_s$ and the modified SPM has to be applied which induces an additional pre-exponential factor 
\begin{eqnarray}
  \frac{1}{[\mathbf{p}(\eta_s)^2+\kappa^2]^2}&=&\frac{1}{4\left(\mathbf{p}(\eta_s)\cdot\dot{\mathbf{p}}(\eta_s)\right)^2(\eta-\eta_s)^2}.
\end{eqnarray} 
The total pre-exponential factor after the $\eta$-integration in Eq.~(\ref{M_rel1}) becomes:
\begin{eqnarray}
  \hat{m}_{fi}(\eta_s)=\frac{2^{\frac{3}{2}-\frac{2I_p}{c^2}}(2I_p)^{\frac{3}{2}-\frac{3I_p}{2c^2}}\exp\left[\frac{2I_p}{c^2}\right]}{\sqrt{\pi}\sqrt{\Gamma\left(3-\frac{2I_p}{c^2}\right)}|E(\eta_s)|^{\frac{3}{2}-\frac{I_p}{c^2}}} \, \hat{m}_{fi},
\label{m_fi3}
\end{eqnarray}
with
\begin{eqnarray}
\hat{m}_{++}
&=&\frac{\sqrt{2}c(2 c+i p_E)\left(1+\frac{\beta}{2}\right)}{ \sqrt{8 c^4+6 c^2 p_E^2+p_E^4}}\\
&&+\frac{\beta ^2 c \left(20 c^4+2 i c^3 p_E+13 c^2 p_E^2+i c p_E^3+2 p_E^4\right)}{3 \sqrt{2}
   (2 c-i p_E) \left(2 c^2+p_E^2\right)^{3/2} \sqrt{4 c^2+p_E^2}},\nonumber\\
\hat{m}_{--}
&=&\frac{\sqrt{2}c(2 c-i p_E)\left(1-\frac{\beta}{2}\right)}{ \sqrt{8 c^4+6 c^2 p_E^2+p_E^4}}\\
&&+\frac{\beta ^2 c \left(20 c^4-2 i c^3 p_E+13 c^2 p_E^2-i c p_E^3+2 p_E^4\right)}{3 \sqrt{2}
   (2 c+i p_E) \left(2 c^2+p_E^2\right)^{3/2} \sqrt{4 c^2+p_E^2}},\nonumber
\end{eqnarray}
and $\hat{m}_{-+}=\hat{m}_{+-}=0$.
In the last step the following expansion is used  
\begin{eqnarray}
  \frac{-\sqrt{2\pi i  \ddot{\tilde{S}}^0(\eta_s)}}{4\left(\mathbf{p}(\eta_s)\cdot\dot{\mathbf{p}}(\eta_s)\right)^2}
  \approx\frac{\sqrt{2\pi}}{4 (2I_p)^{3/4} |E(\eta_s)|^{3/2}}\left(1+\frac{41I_p}{24c^2}\right).
  \label{SPf}
\end{eqnarray}

\begin{figure}
\centering
\includegraphics[width=0.25\textwidth]{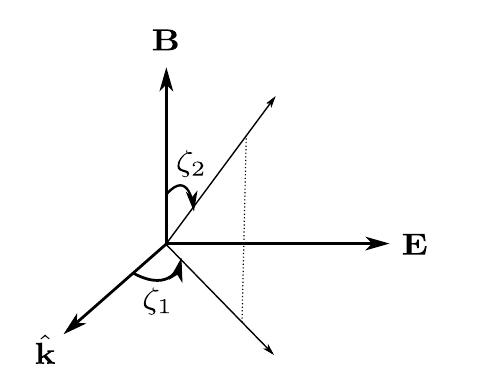}
\caption{Any quantization axis can be aligned with two angles.}
 \label{arb_qa}
\end{figure}

We note that an arbitrary configuration of the quantization axis can be
accomplished via a rotation. Explicitly, since both the initial and
the final state have two independent solutions "up" and "down", i.e., $m_j = \pm 1/2$ for $j=1/2$, the 
rotated state is given by
\be
\ket{1/2,m}' = \sum_{m'} \ket{1/2,m'} D_{m',m}(R)
\ee
where $D_{m',m}(R)$ is the $j=1/2$ representation of the rotation matrix. 
Since the initial quantization axis is along the laser magnetic field, any quantization axis can be aligned with two angles, see Fig. \ref{arb_qa},
with the rotation matrix 
\be
D_{m' m} = \begin{pmatrix}
  e^{-i\zeta_1/2} \cos(\zeta_2/2) & -e^{-i\zeta_1/2}
\sin(\zeta_2/2) \\
& \\
e^{i\zeta_1/2} \sin(\zeta_2/2) & e^{i\zeta_1/2}
\cos(\zeta_2/2)
 \end{pmatrix}.
\ee
Therefore the transition matrix element $\hat{m}'_{fi}$ for arbitrarily rotated spinors is given by
\be
\hat{m}'_{fi} = D^{*}_{f j}\hat{m}_{j k}D_{k i}
\ee
in terms of the initial one $\hat{m}_{fi}$. 
For instance, in the case the quantization axis is along the laser electric field the matrix is
\be
D_{m' m} = \begin{pmatrix}
  \frac{1-i}{2} & -\frac{1-i}{2} \\
& \\
\frac{1+i}{2} & \frac{1+i}{2}
 \end{pmatrix},
\ee
while in the case the quantization axis lays along the laser
propagation direction, it yields
\be
D_{f i} = \begin{pmatrix}
  \frac{1}{\sqrt{2}} & -\frac{1}{\sqrt{2}} \\
& \\
\frac{1}{\sqrt{2}} & \frac{1}{\sqrt{2}}
 \end{pmatrix},
\ee
and the calculation of the corresponding transition amplitudes is straightforward.

\begin{figure}
  \begin{center}
    \includegraphics[width=0.33\textwidth]{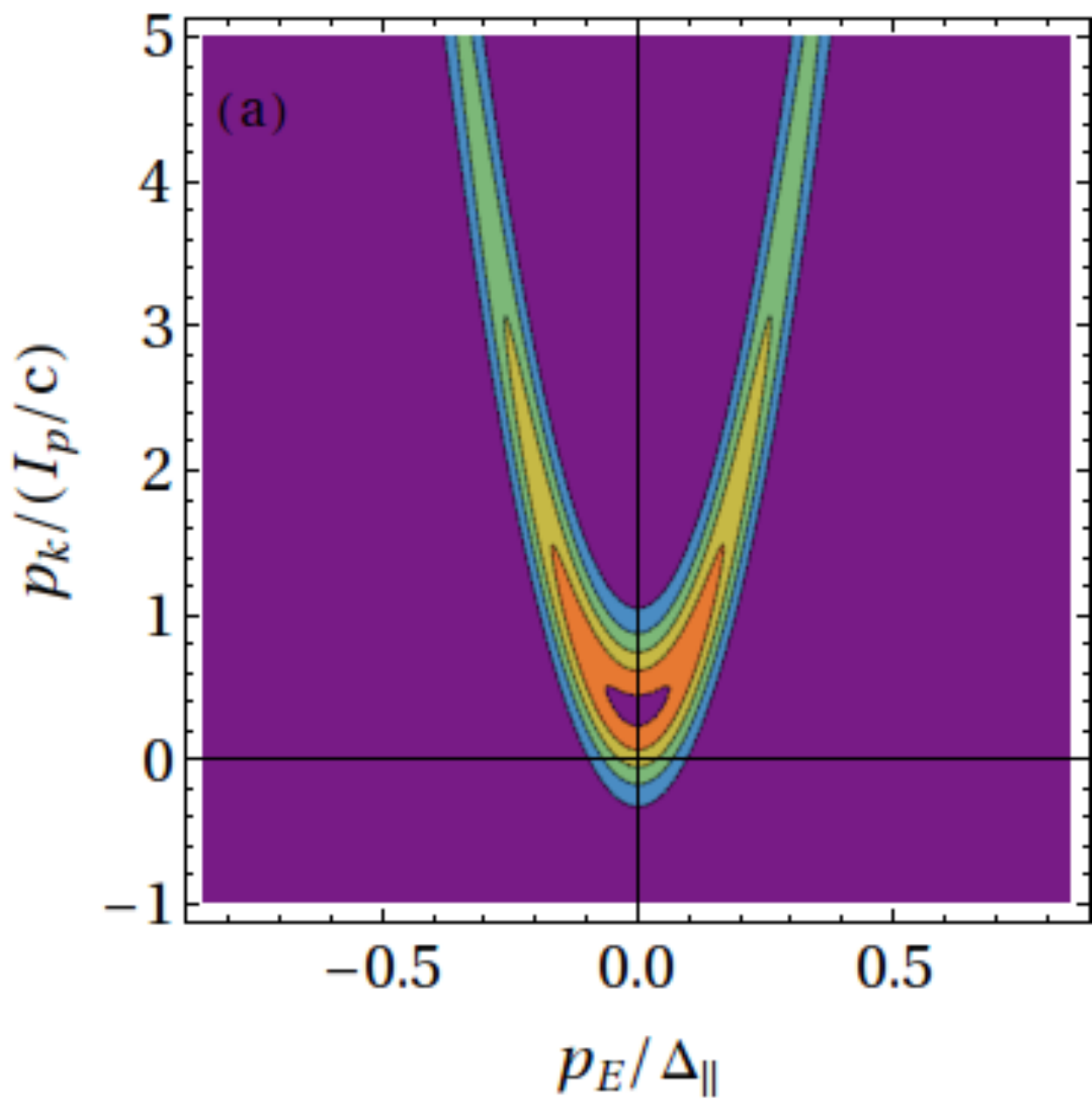}
    \includegraphics[width=0.33\textwidth]{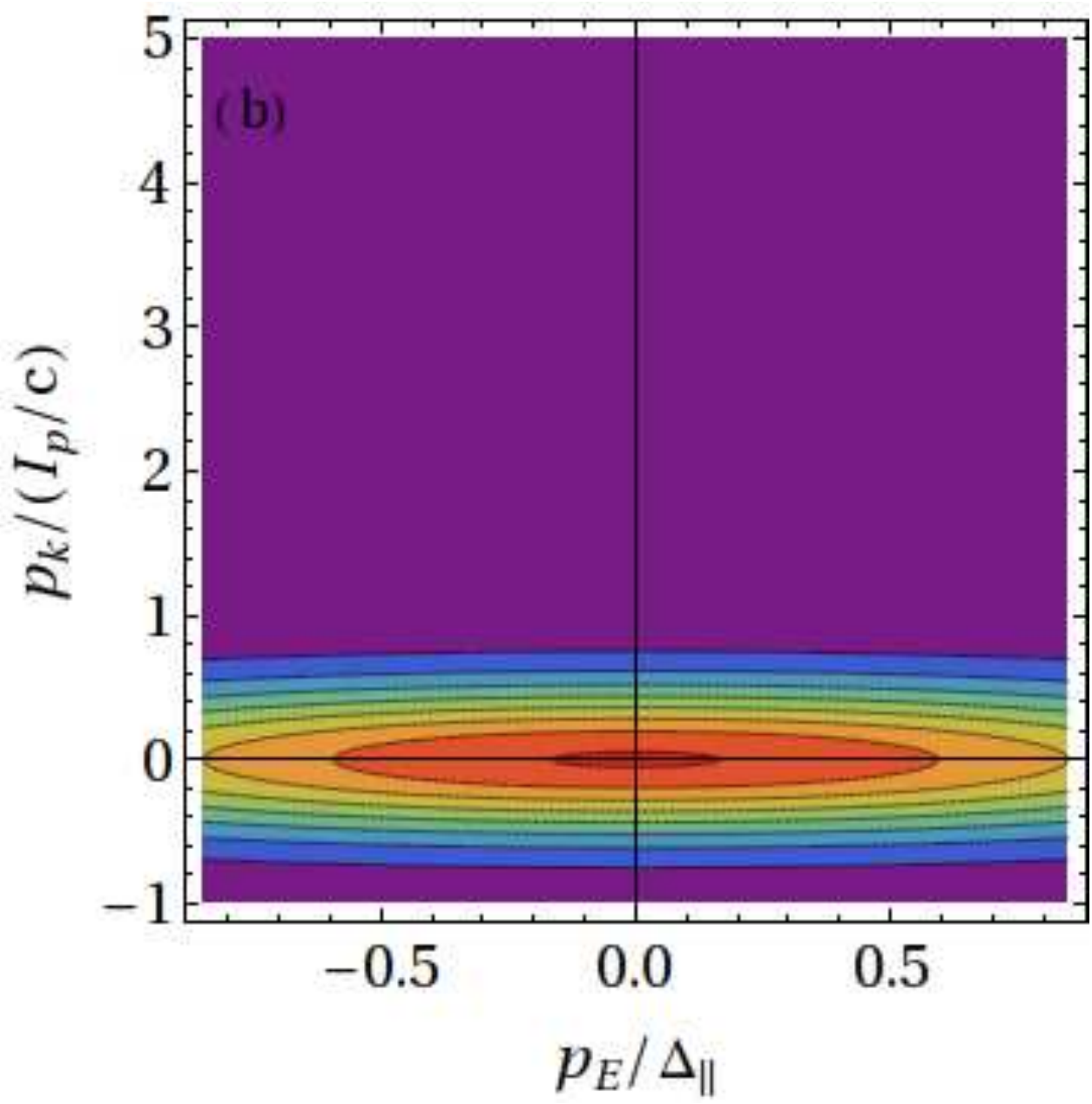}
       \caption{The differential ionization rate for the parameters $I_p/c^2=0.25$, and $E_0/E_a=1/25$: 
(a) relativistic calculation via Eq. (\ref{dwdprela}); (b) non-relativistic calculation via Eq. (I.31), $\Delta_{\parallel}\equiv \sqrt{E_a/E_0}(E_0/\omega)$ is the longitudinal momentum width.}
    \label{dwdp}
  \end{center}
\end{figure}

The differential ionization rate averaged by the initial spin polarization and summed up over the final polarizations is defined as
\begin{eqnarray}
  \frac{dw}{d^3\mathbf{p}}&=&\frac{\omega}{2\pi}\Big(|\hat{m}_{++}(\eta_s)|^2+|\hat{m}_{+-}(\eta_s)|^2+|\hat{m}_{-+}(\eta_s)|^2\nonumber\\&&
  +|\hat{m}_{--}(\eta_s)|^2\Big)\exp\{-2\,{\rm Im}\,[S^0(\eta_s)]\},
  \label{dwdprela0}
\end{eqnarray}
which with the help of Eq. (\ref{m_fi3}) will read
\begin{eqnarray}
    \frac{dw}{d^3\mathbf{p}}&=&\frac{ 2^{3-\frac{4I_p}{c^2}}\left(1+\frac{p_E^2}{2c^2}-\frac{5I_p}{4c^2}-\frac{19p_E^2I_p}{24c^4}\right)(2I_p)^{3\left(1-\frac{I_p}{c^2}\right)}\omega\exp\left(\frac{4I_p}{c^2}\right)}{\pi^2\Gamma\left(3-\frac{2I_p}{c^2}\right)|E(\eta_s)|^{3-\frac{2I_p}{c^2}}\left(1+\frac{p_E^2}{2c^2}\right)^2}\nonumber\\
  &&\times\left(1+\frac{41I_p}{12c^2}\right)\exp\left\{\frac{-2\left[2\Lambda(\varepsilon-c^2+I_p)-p_E^2\right]^{3/2}}{3|\mathbf{E}(\eta_s)|\Lambda}\right\}.\nonumber\\
  \label{dwdprela}
\end{eqnarray}
Note that there is an additional factor of two, that arises due to summing up over the two saddle points per laser cycle. The momentum distribution at ionization in the relativistic regime is plotted in Fig.~\ref{dwdp}. As it is already mentioned, the distribution is located around a parabola, see also \cite{Krainov_1992,Krainov_1999,Krainov_2003,Krainov_2004,Krainov_2008}.

Characteristic features of the momentum distribution in the relativistic regime are the large parabolic wings corresponding to the large longitudinal momentum (along the laser propagation direction) which arise during the free electron motion in the laser field. However, a more interesting feature of this distribution is the small shift of the peak of the distribution with respect to the nonrelativistic one. While the nonrelativistic distribution peak is at $p_E=0, \,p_k=0$, in the relativistic case it is shifted to $p_E=0, \,p_k=I_p/3c$. As we have shown in \cite{Klaiber_2012}, this momentum shift is due to the under-the barrier dynamics and arises already at the ionization tunnel exit but, notwithstanding of the large momentum transfer from the laser field during the electron excursion after the ionization, the characteristic momentum shift of the peak of the distribution is detectable far away from the interaction zone  at the detector.

When the exponent in the differential ionization rate is expanded quadratically around the parabola $p_E=p_E$, $p_B=\delta p_B$, $p_k=I_p/3c+p_E^2/2c(1+I_p/3c^2)+\delta p_k$, the rate expression reads
\begin{eqnarray}
  \frac{dw}{d^3\mathbf{p}}&=&\frac{ 2^{3-\frac{4I_p}{c^2}}\left(1+\frac{p_E^2}{2c^2}-\frac{5I_p}{4c^2}-\frac{19p_E^2I_p}{24c^4}\right)\exp\left(\frac{4I_p}{c^2}\right) (2I_p)^{3\left(1- \frac{I_p}{c^2}\right)}\omega}{\pi^2\Gamma\left(3-\frac{2I_p}{c^2}\right)|E(\eta_s)|^{3-\frac{2I_p}{c^2}}\left(1+\frac{p_E^2}{2c^2}\right)^2}\nonumber\\
  &&\times\left(1+\frac{41I_p}{12c^2}\right)\exp\left\{-\frac{2E_a}{3|E(\eta_s)|}\left(1-\frac{I_p}{12c^2}\right)\right.\\ 
  &&\left.-\frac{\sqrt{2I_p}}{|E(\eta_s)|}\left[\frac{\delta p_B^2}{\left(1 - \frac{7I_p}{72c^2}\right)^2}+\frac{\delta p_k^2}{\left(1 + \frac{I_p}{8c^2}+ \frac{p_E^2}{2c^2}\left(1+\frac{11I_p}{24c^2}\right)\right)^2}\right]\right\}.\nonumber
\label{dwdprel2}
\end{eqnarray} 
In Eq. (\ref{dwdprela}) an expansion on the $I_p/c^2$ parameter has been used to clearly indicate the $I_p/c^2$-scaling. Without $I_p/c^2$-expansion the exponential factor of the differential ionization rate reads 
\begin{eqnarray}
\frac{dw}{d^3\mathbf{p}}&\propto&\exp\left\{-\frac{2 \sqrt{3}\Xi^3c^3}{(1 + \Xi^2)|E(\eta_s)|}-\frac{\sqrt{I_p}}{E(\eta_s)}\frac{\sqrt{3}\Xi\,\,\delta p_B^2}{\sqrt{1 +\frac{  (\Xi^2-1)}{\sqrt{1 + \Xi^2}}}}\right.\nonumber\\
&&-\left. \frac{\sqrt{2I_p}}{E(\eta_s)}\sqrt{\frac{8}{3}}\Xi 
   \left(\Xi ^2+3\right) \frac{\delta p_k^2}{{\cal D}^2}\right\}, 
\end{eqnarray}
where $\Xi=\sqrt{1-\Upsilon /2(\sqrt{\Upsilon^2+8}-\Upsilon)}$, $\Upsilon=1-I_p/c^2$, and
\begin{eqnarray}
{\cal D}&\equiv&  \left(\Xi ^2+2\right) \sqrt[4]{\sqrt{\Xi ^2+1}-\frac{2}{\sqrt{\Xi ^2+1}}+1}\\
&+&\frac{p_E^2}{c^2}\sqrt[4]{\left(\Xi ^2+1\right)^3 \left(\left(\sqrt{\Xi ^2+1}+1\right) \Xi ^2-\sqrt{\Xi ^2+1}+1\right)}.\nonumber
\end{eqnarray}
Using the expression Eq. (\ref{dwdprel2}), the integration over momentum space can be done analytically. It yields for the total ionization rate:
\begin{eqnarray}
  w&=&\frac{2^{3-\frac{4I_p}{c^2}}}{\Gamma\left(3-\frac{2I_p}{c^2}\right)}\sqrt{\frac{3}{\pi}}\left(1+\frac{161I_p}{72c^2}\right)\exp\left(\frac{4I_p}{c^2}\right)\nonumber\\
  &&\times\frac{(2I_p)^{\frac{7}{4}-\frac{3I_p}{c^2}}}{E_0^{\frac{1}{2}-\frac{2I_p}{c^2}}}\exp\left[-\frac{2E_a}{3E_0}\left(1-\frac{I_p}{12c^2}\right)\right].
  \label{ADKrel}
\end{eqnarray}
\begin{figure}
  \begin{center}
    \includegraphics[width=0.4\textwidth]{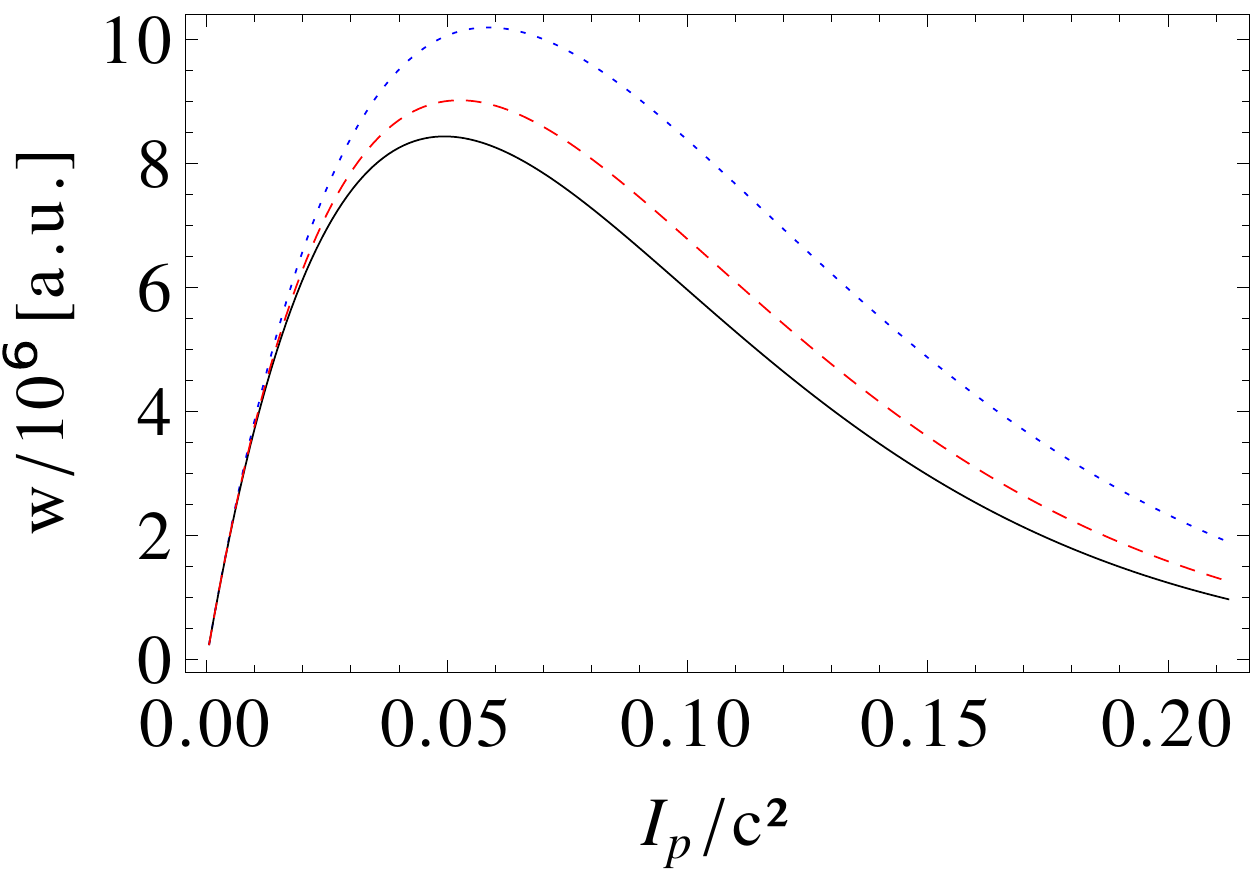}
        \caption{The total ionization rate vs ionization potential for $E_0/E_a=1/32$: (blue, short-dashed) in the relativistic Coulomb-corrected SFA via Eq.~(\ref{ADKrel}), (red, long-dashed) in the relativistic Coulomb-corrected modified SFA via Eq.~(\ref{ADKrel2}), and (black, solid) relativistic PPT from Ref.~\cite{Milosevic_2002r2}.}
    \label{WV}
  \end{center}
\end{figure}
In Fig. \ref{WV} the total ionization rate of Eq. (\ref{ADKrel}) is compared with the ITM result of Ref.\cite{Milosevic_2002r2}. The Coulomb-corrected relativistic SFA and the relativistic PPT yields results for the total ionization rate which are close, though they are not identical. In the next section we  modify the Coulomb-corrected SFA with the aim to obtain ionization probabilities closer to the relativistic PPT result.

\section{Modified Coulomb-corrected relativistic strong-field approximation}\label{modified}

In the standard SFA in the G\"oppert-Mayer gauge, the transition matrix element is given by Eq. (\ref{M_rel}) where 
\begin{eqnarray}
H_{int}=\mathbf{r}\cdot\mathbf{E}(1-\hat{\mathbf{k}}\cdot\boldsymbol{\alpha})
\end{eqnarray}
and the wave function of the initial state $\phi^{(c)}_{i}(\mathbf{r},t)$ describes the free evolving atomic system, fulfilling the equation 
\begin{eqnarray}
(i\partial_t-H_0 )\phi^{(c)}_{i}(\mathbf{r},t)=0, 
\end{eqnarray}
with $H_0=H-H_{int}= H_0=c\boldsymbol{\alpha}\cdot\hat{\mathbf{p}}+\beta c^2+V^{(c)}(\mathbf{r})$. However, in general, as it is pointed out in \cite{Faisal_2007a,Faisal_2007b},  the SFA can be developed employing different partitions of the total Hamiltonian which will result in different physical approximations. The main deficiency of the Coulomb-corrected SFA based on the standard partition indicated above, is that the influence of the laser field on the atomic bound dynamics is completely neglected. In this section we propose a modified version of the SFA employing another partition of the Hamiltonian in the SFA formalism. The main motivation is to use such a partition where the laser field has a contribution to the bound state evolution. In particular, we use the following partition:
\begin{eqnarray}
  H\,\,\,&=&\tilde{H}_0+\tilde{H}_{int},\\
\tilde{H}_0\,\, &=& c\boldsymbol{\alpha}\cdot\left[ \hat{\mathbf{p}}-\hat{\textbf{k}}\left(\textbf{r}\cdot \textbf{E}(\eta)\right)/c\right]+\beta c^2+V^{(c)}(r)\\
\tilde{H}_{int} &=& \mathbf{r}\cdot\mathbf{E}(\eta).
\end{eqnarray}
In this partition the transition matrix element reads
\begin{eqnarray}
  \tilde{M}^{(c)}_{fi}=-i\int dt d^3\mathbf{r}\psi^+_{f}(\mathbf{r},t)\tilde{H}_{int}\tilde{\phi}^{(c)}_{i}(\mathbf{r},t),
  \label{M_rel2}
\end{eqnarray}
where the bound state dynamics is determined by the following time-dependent Dirac equation: 
\begin{eqnarray}
  i\partial_t\tilde{\phi}^{(c)}_{i}=\left\{c\boldsymbol{\alpha}\cdot\left[ \hat{\mathbf{p}}-\hat{\textbf{k}}\left(\textbf{r}\cdot \textbf{E}(\eta)\right)/c\right]+\beta c^2+V^{(c)}(r) \right\}\tilde{\phi}^{(c)}_{i}.
\label{modified_eq_i}
\end{eqnarray}
In this specific partition the initial bound state wave function $\tilde{\phi}^{(c)}_{i}$ is the eigenstate of the instantaneous energy operator 
\begin{eqnarray}
\hat{{\cal E}}=c\boldsymbol{\alpha}(\mathbf{p}+\mathbf{A}(\eta) )+\beta c^2
\end{eqnarray}
\cite{Scully_Zubairy_1997,Klaiber_2006a}, with $\mathbf{A}(\eta)=-\hat{\textbf{k}} (\textbf{r}\cdot\textbf{E}(\eta))/c$ in the G\"oppert-Mayer gauge, see Eq. (\ref{A_GM}). Note that in the nonrelativistic limit the standard SFA in the length gauge corresponds to the partition in which the initial bound state is the eigenstate of the energy operator \cite{Faisal_2007a}. This point provides another argument in favor of the  applied partition in the relativistic case.

The term incorporating the laser electric field in Eq. (\ref{modified_eq_i}) describes the spin dynamics of the bound electron before ionization and has to be considered in the further calculation. 
We are concerned by the spin dynamics in the bound state and by the dynamical Zeeman splitting of the bound state energies due to the spin interaction with the laser magnetic field. Therefore, in solving Eq. (\ref{modified_eq_i}) we will restrict the bound-bound transitions only to the transitions between different spin-states at  fixed quantum numbers $\{n,j,m_j\}$. The dipole approximation for the laser field will be adopted for the description of the bound state evolution: $\textbf{E}(\eta)\approx \textbf{E}(t)$, since the typical length-scale for the bound dynamics is much smaller than the laser wave length: $r_b/\lambda\sim\omega/\kappa c\sim\gamma (E_0/E_a)(\kappa/c)\ll 1$. When the initial spin-polarization is along the laser magnetic field $|\phi^{(c)} \rangle=|\phi^{(c)}_{B,\pm}\rangle$, then no spin transitions occur because the interaction term 
$\hat{V}_{\rm int}(t)\equiv -(\boldsymbol{\alpha}\cdot\hat{\textbf{k}})(\textbf{r}\cdot \textbf{E}(t))$ does not cause  spin-flip  $\langle \phi^{(c)}_{B,\mp}|\hat{V}_{\rm int}(t)|\phi^{(c)}_{B,\pm}\rangle=0$. Accordingly, in this case we are looking for the solution of Eq.~(\ref{modified_eq_i}) in the form: 
\begin{eqnarray}
   \tilde{\phi}^{(c)}_{B,\pm}(\textbf{r},t)= \phi^{(c)}_{B,\pm}(\textbf{r},t)e^{i S(t)},
\label{solution_form}
\end{eqnarray}
where $\phi^{(c)}_{B,\pm}(\textbf{r},t)=\phi^{(c)}_{B,\pm}(\textbf{r})e^{-i\varepsilon_{\pm}t }$ is the ground state before switching on the laser field, i.e., an eigenstate of the atomic unperturbed Hamiltonian
 \begin{eqnarray}
  \left[c\boldsymbol{\alpha}\cdot \hat{\mathbf{p}}+\beta c^2+V^{(c)}(r)\right]\phi^{(c)}_{B,\pm}(\textbf{r})=\varepsilon_{\pm}\phi^{(c)}_{B,\pm}(\textbf{r}).
 \end{eqnarray}
The spin quantization axis for $\phi^{(c)}_{B,\pm}(\textbf{r},t)$ states is along the laser magnetic field direction. 
Taking into account that 
\begin{eqnarray}
 \int_{-\infty}^\eta dt' \langle \phi^{(c)}_{B,\pm}|\hat{V}_{\rm int}(t')|\phi^{(c)}_{B,\pm}\rangle=\frac{\tilde{A}(\eta)}{2c} \left(1 - \frac{2 I_p}{3c^2}\right), 
\end{eqnarray}
where $\tilde{A}(\eta)\equiv -\int_{-\infty}^\eta E(\eta') d\eta'$, the wave function of the bound state will read
\begin{eqnarray}
  \tilde{\phi}^{(c)}_{B,+}(t)&=&\phi^{(c)}_{B,+}(t)\exp\left[i\frac{\tilde{A}(\eta)}{2c} \left(1 - \frac{2 I_p}{3c^2}\right)\right]\nonumber\\
  \tilde{\phi}^{(c)}_{B,-}(t)&=&\phi^{(c)}_{B,-}(t)\exp\left[-i\frac{\tilde{A}(\eta)}{2c} \left(1 - \frac{2 I_p}{3c^2}\right)\right]
\label{TE}
\end{eqnarray}
The Zeeman splitting term in the phase of the wave function is of order of $\tilde{A}/c\sim E_0\tau_c/c\sim \kappa/c\ll 1$. It is small which arises from the fact that the perturbation term $\hat{V}_{\rm int}(t)$ couples the large/small  spinor part of the initial state bispinor with the small/large spinor parts of the final state bispinor.

The explicit condition justifying the neglect of transitions to excited states can be given as follows. The transition probability between the states $n\rightarrow n^\prime$ with energy difference $\omega_{n\,n^\prime}\sim I_p$ ($n$ and $n^\prime$ are the principal quantum numbers) is $P_{n\,n^\prime}\sim E_0 r_b/\omega_{n\,n^\prime}\sim E_0 r_b/I_p$, while the probability of a spin-transition in the state $n$ is $P_{s\,s^\prime}\sim E_0 r_b \tau_K$  ($s$ and $s^\prime$ are the spin quantum numbers). Therefore,
\begin{eqnarray}
\frac{P_{n\,n^\prime}}{P_{s\,s^\prime}  }\sim \frac{1}{I_p\tau_K }\sim \frac{E_0}{E_a}, 
  \end{eqnarray}
and the  transitions to excited states are suppressed by a factor of $E_0/E_a$.

With the wave functions of the bound state Eq.~(\ref{TE}), the differential and the total ionization rates can be calculated in the same way as in the previous section before. The structure of the ionization matrix element remains the same
\begin{eqnarray}
M^{(c)}_{fi}=-i\tilde{m}_{fi}(\eta_s)\exp\left\{-\frac{\left[2\Lambda(\varepsilon-c^2+I_p)-p_E^2\right]^{3/2}}{3|\mathbf{E}(\eta_0)|\Lambda}\right\}
\end{eqnarray}
where $\tilde{m}_{fi}(\eta_s)$ is given by Eq.~(\ref{m_fi11}), in which the following replacement should be done in order to  account for the  difference in the spinor operator of the interaction Hamiltonian in the modified SFA:
\begin{eqnarray}
  u_f^+\left(1-\hat{\mathbf{k}}\cdot\boldsymbol{\alpha}\right)v_i\rightarrow u_f^+\left[1+\frac{\mathbf{A}(\eta_s)\cdot\boldsymbol{\alpha}(1+\hat{\mathbf{k}}\cdot\boldsymbol{\alpha})}{2c\Lambda}\right]\tilde{v}_i,
\label{modified_spin}
\end{eqnarray}
where $\tilde{v}_i=v_i\exp[\pm i\tilde{A}(\eta)(1-2I_p/3c^2)]$ is the spinor part of the states given in Eq.~(\ref{TE}). The spinor operator in brackets in Eq. (\ref{modified_spin})  comes from the spinor part of the Volkov wave function. Note that the phase in the exponential in Eq.~(\ref{TE}) varies slowly compared with the contracted action $\tilde{S}$. Therefore, it does not modify the saddle point integration, but alters the  pre-exponential term. 
\begin{eqnarray}
&&\tilde{m}_{++}=\frac{\sqrt{2} c e^{-\frac{i p_E}{2 c}} (2 c+i p_E)}{\sqrt{8 c^4+6 c^2 p_E^2+p_E^4}}\\
&&-\frac{\beta ^2 e^{-\frac{i
   p_E}{2 c}} (2 c+i p_E) \left(8 c^5-40 i c^4 p_E+14 c^3 p_E^2\right)}{12 \sqrt{2} \left(8 c^4+6 c^2 p_E^2+p_E^4\right)^{3/2}},\nonumber\\
&&-\frac{\beta ^2 e^{-\frac{i
   p_E}{2 c}} (2 c+i p_E) \left(-28 i c^2 p_E^3+3 c p_E^4-4 i
   p_E^5\right)}{12 \sqrt{2} \left(8 c^4+6 c^2 p_E^2+p_E^4\right)^{3/2}},\nonumber
\end{eqnarray}
\begin{eqnarray}
&&\tilde{m}_{--}=\frac{\sqrt{2} c e^{\frac{i p_E}{2 c}} (2 c-i p_E)}{\sqrt{8 c^4+6 c^2 p_E^2+p_E^4}}\\
&&-\frac{\beta ^2 e^{\frac{i
   p_E}{2 c}} \left(8 c^5+40 i c^4 p_E+14 c^3 p_E^2+28 i c^2 p_E^3+3 c p_E^4+4 i p_E^5\right)}{12 \sqrt{2} (2
   c+i p_E) \left(2 c^2+p_E^2\right) \sqrt{8 c^4+6 c^2 p_E^2+p_E^4}}\nonumber
\end{eqnarray}
and $\tilde{m}_{+-}=\tilde{m}_{-+}=0$.
With this the differential ionization rate in the modified SFA yields:
\begin{eqnarray}
    \frac{dw}{d^3\mathbf{p}}&=&\frac{ 2^{3-\frac{4I_p}{c^2}}\left(1+\frac{p_E^2}{2c^2}-\frac{37I_p}{12c^2}-\frac{41p_E^2I_p}{24c^4}\right)(2I_p)^{3\left(1-\frac{I_p}{c^2}\right)}\omega\exp\left(\frac{4I_p}{c^2}\right)}{\pi^2\Gamma\left(3-\frac{2I_p}{c^2}\right)|E(\eta_s)|^{3-\frac{2I_p}{c^2}}\left(1+\frac{p_E^2}{2c^2}\right)^2}\nonumber\\
  &\times&\left(1+\frac{41I_p}{12c^2}\right)\exp\left[\frac{-2\left(2\Lambda(\varepsilon-c^2+I_p)-p_E^2\right)^{3/2}}{3|\mathbf{E}(\eta_s)|\Lambda}\right].
  \label{dwdprela1}
\end{eqnarray}
and the total ionization rate is
\begin{eqnarray}
  w&=&\frac{2^{3-\frac{4I_p}{c^2}}}{\Gamma\left(3-\frac{2I_p}{c^2}\right)}\sqrt{\frac{3}{\pi}}\left(1-\frac{7I_p}{72c^2}\right)\exp\left(\frac{4I_p}{c^2}\right)\\
  &&\times\frac{(2I_p)^{\frac{7}{4}-\frac{3I_p}{c^2}}}{E_0^{\frac{1}{2}-\frac{2I_p}{c^2}}}\exp\left[-\frac{2E_a}{3E_0}\left(1-\frac{I_p}{12c^2}\right)\right].\nonumber
  \label{ADKrel2}
\end{eqnarray}

In Fig.~\ref{WV}, the total ionization probability  calculated via the modified Coulomb-corrected relativistic SFA is compared with the result of the  standard SFA, as well as with that of the relativistic PPT.  The conclusion can be drawn that both relativistic Coulomb-corrected SFA give slightly different results compared to the PPT for large $I_p$. The modified SFA is closer to the PPT result  than the standard SFA.  The total ionization probabilities in the relativistic Coulomb-corrected standard SFA is larger than the relativistic PPT by a factor smaller than $1+3I_p/c^2$. 

The standard SFA yields larger ionization probabilities than the modified SFA which can be explained by a simple tunneling picture. In the first case the magnetic field  acts on the electron  only when it enters the barrier. In this case during the tunneling the electron kinetic energy will change on the amount of the interaction energy with the magnetic field $\mu B_0=-s E_0/2c$ ($s=\pm 1$ for spin along the magnetic field or opposite), giving rise to a kinetic energy splitting for different spin orientations of the electron and, consequently, to different tunneling probabilities.
Whereas, in the second case the magnetic field is also acting before tunneling and no kinetic energy splitting occurs. The ratios of the Keldysh-exponent for these two scenarios can then be expanded in the spin energy:
\begin{eqnarray}
  \frac{\Gamma_{\rm standard}}{\Gamma_{\rm modified}}&=&\frac{\exp\left[-\frac{2\left(2(I_p+E_0/2c)\right)^{3/2}}{3E_0}\right]+\exp\left[-\frac{2\left(2(I_p-E_0/2c)\right)^{3/2}}{3E_0}\right]}{2\exp\left[-\frac{2\left(2(I_p)\right)^{3/2}}{3E_0}\right]}\nonumber\\
&\sim&1+I_p/c^2
\end{eqnarray}
To decide which SFA partition is best suited to model the ionization process, a comparison with numerical simulations or experimental data is necessary.

Further, we note that the Coulomb-correction term $\int d\eta \boldsymbol{\alpha}\cdot\boldsymbol{\nabla}V^{(c)}/c$ that is neglected in both strong field approximations 
gives a correction factor of only $1+(I_p/c^2)\sqrt{E_0/E_a}$, which is of the order of a few percents for the most extreme parameters ($E_0/E_a=1/16$, $Z=90$).

\section{Conclusion}

We have generalized the Coulomb-corrected SFA for the ionization of hydrogen-like systems in a strong laser field into the relativistic regime. The applied Coulomb-corrected strong-field approximation incorporates the eikonal-Volkov wave function for the description of the electron continuum dynamics. The latter is derived in the WKB approximation taking into account the Coulomb field of the atomic core perturbatively in the phase of the WKB wave function. In physical terms, the disturbance of the electron energy by the Coulomb field is assumed to be smaller with respect to the electron energy in the laser field.
The eikonal-Volkov wave function is applicable when the laser field is smaller than the atomic field $E_0\ll E_a$ and $I_p\lesssim c^2$.

We have derived an analytical expression for the ionization amplitude in a linearly polarized laser field within the Coulomb-corrected relativistic SFA when additionally  smallness of the parameters of $I_p/c^2\ll $ and $\gamma\ll 1$ is used. A simple expression for the amplitude is obtained when using the G\"oppert-Mayer gauge. Moreover, in this gauge a Coulomb correction factor (ratio of the Coulomb 
corrected amplitude to the standard SFA one) coincides with that derived within the PPT theory. The differential and total ionization rates are calculated analytically. The calculated total ionization rate is slightly larger than the PPT rate at large ionization potentials. To improve the predictions of the Coulomb-corrected relativistic SFA, we have proposed a modified SFA approach which is based on another partition of the total Hamiltonian. In this approach the SFA matrix element contains the eigenstate of the energy operator in the laser field as the wave function of the initial bound state. The modified SFA takes into account the dynamical Zeeman splitting of the bound state energy due to the spin interaction with the laser magnetic field (when the electron initial polarization is along the laser magnetic field) and the precession of the electron spin in the bound state (when the electron initial polarization is along the laser propagation direction).

Our results show that the SFA technique allows the analytical calculation of quantitatively correct differential and total ionization rates in the relativistic regime which takes into account the impact of the Coulomb field of the atomic core as well as the electron spin dynamics in the bound state. This method can be viewed as an alternative to the PPT. In the following paper of the sequel the Coulomb-corrected relativistic SFA will be used to investigate spin-resolved ionization probabilities. While for the total ionization rate the prediction of the standard SFA is close to that of the modified SFA, for spin effects their predictions are quite different. The next paper of the sequel will be devoted to this issue.

\section*{Acknowledgments}

We appreciate valuable discussions with C. H. Keitel and C. M\"uller.

\bibliography{strong_fields_bibliography}

\end{document}